\documentclass[sigconf,anonymous=False]{acmart}

\AtBeginDocument{%
  \providecommand\BibTeX{{%
    \normalfont B\kern-0.5em{\scshape i\kern-0.25em b}\kern-0.8em\TeX}}}
\usepackage[labelformat=simple]{subcaption}

\usepackage{enumitem}
\setlist{nosep}
\usepackage{bm}
\usepackage[linesnumbered,ruled]{algorithm2e}
\let\oldnl\nl
\newcommand{\nonl}{\renewcommand{\nl}{\let\nl\oldnl}}

\usepackage{algpseudocode}

\usepackage{booktabs}
\usepackage{float}
\usepackage{overpic}
\usepackage{float}  
\usepackage{subfloat}
\usepackage{bm}
\usepackage{multirow}
\usepackage{graphicx}
\usepackage{subcaption}
\usepackage{colortbl}
\usepackage{enumitem}
\usepackage{stfloats} 
\usepackage{url}
\usepackage{verbatim}
\usepackage{collab}
\usepackage[normalem]{ulem}
\useunder{\uline}{\ul}{}
\usepackage[skip=0.5\baselineskip]{caption}
\usepackage{color, xspace}
\usepackage{tcolorbox}  
\tcbuselibrary{breakable}   

\newcommand{\zxy}[1]{{\color{blue} [xiangyu: #1]}}

\newcommand{\eat}[1]{}
\definecolor{editBlue}{RGB}{134,150,167}

\copyrightyear{2024}
\acmYear{2024}
\setcopyright{acmlicensed}
\acmConference[CIKM '24]{Proceedings of the 33rd ACM International Conference on Information and Knowledge Management}{October 21--25, 2024}{Boise, ID, USA}
\acmBooktitle{Proceedings of the 33rd ACM International Conference on Information and Knowledge Management (CIKM '24), October 21--25, 2024, Boise, ID, USA}
\acmDOI{10.1145/3627673.3679743}
\acmISBN{979-8-4007-0436-9/24/10}

\usepackage{xspace}

\newcommand{\name}{LLM4MSR\xspace}

\usepackage{marvosym}
\usepackage{ifsym}

\author{Yuhao Wang}
\affiliation{%
  \institution{City University of Hong Kong}
  \city{Hong Kong}
  \country{China}
}
\email{yhwang25-c@my.cityu.edu.hk}

\author{Yichao Wang}
\affiliation{%
  \institution{Huawei Noah’s Ark Lab}
  \city{Shenzhen}
  \country{China}
}
\email{wangyichao5@huawei.com}

\author{Zichuan Fu}
\affiliation{%
  \institution{City University of Hong Kong}
  \city{Hong Kong}
  \country{China}
}
\email{zichuanfu2-c@my.cityu.edu.hk}

\author{Xiangyang Li}
\affiliation{%
  \institution{Huawei Noah’s Ark Lab}
  \city{Shenzhen}
  \country{China}
}
\email{lixiangyang34@huawei.com}

\author{Wanyu Wang}
\affiliation{%
  \institution{City University of Hong Kong}
  \city{Hong Kong}
  \country{China}
}
\email{wanyuwang4-c@my.cityu.edu.hk}

\author{Yuyang Ye}
\affiliation{%
  \institution{City University of Hong Kong}
  \city{Hong Kong}
  \country{China}
}
\email{yuyangye@cityu.edu.hk}

\author{Xiangyu Zhao \Letter}
\thanks{\Letter \text{Corresponding author}}
\affiliation{%
  \institution{City University of Hong Kong}
  \city{Hong Kong}
  \country{China}
}
\email{xianzhao@cityu.edu.hk}

\author{Huifeng Guo} 
\affiliation{%
  \institution{Huawei Noah’s Ark Lab}
  \city{Shenzhen}
  \country{China}
}
\email{huifeng.guo@huawei.com}

\author{Ruiming Tang}
\affiliation{%
  \institution{Huawei Noah’s Ark Lab}
  \city{Shenzhen}
  \country{China}
}
\email{tangruiming@huawei.com}

\renewcommand{\shortauthors}{Yuhao Wang et al.}

\begin{document}

\title{\name: An LLM-Enhanced Paradigm for Multi-Scenario Recommendation}

\begin{abstract}
As the demand for more personalized recommendation grows and a dramatic boom in commercial scenarios arises, the study on multi-scenario recommendation (MSR) has attracted much attention, which uses the data from all scenarios to simultaneously improve their recommendation performance.
However, existing methods tend to integrate insufficient scenario knowledge and neglect learning personalized cross-scenario preferences, thus leading to sub-optimal performance. 
Meanwhile, though large language model (LLM) has shown great capability of reasoning and capturing semantic information, the high inference latency and high computation cost of tuning hinder its implementation in industrial recommender systems.
To fill these gaps, we propose an 
LLM-enhanced paradigm \name in this work.
Specifically, we first leverage LLM to uncover multi-level knowledge 
from the designed scenario- and user-level prompt without fine-tuning the LLM, then adopt hierarchical meta networks 
to generate multi-level meta layers 
to explicitly improve the scenario-aware and personalized recommendation capability.
Our experiments on KuaiSAR-small, KuaiSAR, and Amazon datasets validate significant advantages of \name: (i) the effectiveness and compatibility with different multi-scenario backbone models,
(ii) high efficiency and deployability on industrial recommender systems, and (iii) improved interpretability.
The implemented code and data is available to ease reproduction\footnote{\label{foot1}\url{https://github.com/mindspore-lab/models/tree/master/research/huawei-noah/LLM4MSR}}$^,$\footnote{\label{foot11}\url{https://github.com/Applied-Machine-Learning-Lab/LLM4MSR}}.
\end{abstract}



\maketitle
\section{Introduction} \label{intro}
The research on recommender system (RS) has been a hot spot in the past few years which targets at mining users' interests from the vast amount of historical interaction data.
In order to provide services in a more personalized manner and make profits, an explosive growth of business scenarios (also known as domains) are emerging in commercial RSs such as the search and recommendation services on mobile apps \cite{gong2023unified,Sun2023KuaiSAR} and different product categories on the e-commerce platform \cite{ni2019justifying}.
Therefore, as a group of efficient joint modeling approach, multi-scenario recommendation (MSR) has drawn attention increasingly, which exploits all data to improve the recommendation accuracy on all these scenarios, thus addressing the data sparsity issue \cite{zhu2021cross} and reducing computation cost.

Nevertheless, existing MSR methods usually suffer from the following two limitations: 
\textbf{(i)} Insufficient scenario knowledge is incorporated. Specifically, domain indicator is usually the only scenario knowledge leveraged to capture the domain distinction in practice \cite{sheng2021one,chang2023pepnet}, while the abundant semantic information such as the description of scenario is neglected. 
Consequently, the correlations among scenarios are not adequately established. 
\textbf{(ii)} Users' personalized preferences across scenarios tend to be ignored, since most MSR models merely rely on different parameter sharing patterns in multi-task learning \cite{sheng2021one,wang2023multi,liu2023multi} and the collaborative signal learned in conventional RSs to conduct recommendation.
However, these deficiencies are still overlooked even if they would lead to the sub-optimal performance of multi-scenario recommender system.

To tackle these drawbacks of conventional MSR models, we resort to large language model (LLM) which is known for its remarkable capability of language understanding and reasoning \cite{fan2023recommender}. Although there have been some attempts to enhance RSs harnessing LLM \cite{gong2023unified,yin2023heterogeneous,li2023ctrl}, there are some key challenges.
The first is high inference latency of LLM since the existing methods conduct training and inference with LLM through sample level alignment \cite{li2023ctrl,wang2023alt}.
Second, the information generated from LLM is usually in natural language format. In other words, it requires filling the gap between semantic and recommendation space like employing additional knowledge encoder \cite{xi2023towards} and instruction tuning on LLM \cite{yin2023heterogeneous} which brings high computation cost. Moreover, both conventional RS and LLM-based recommendation methods lack interpretability.

To this end, we propose an LLM-based paradigm to enhance conventional multi-scenario recommendation, which leverages LLM and hierarchical meta networks to explicitly improve the performance of the backbone model on all scenarios. To be specific, on the one hand, LLM helps grasping the cross-scenario correlation and personalized preferences, while the meta networks act as a flexible and adaptive bridge connecting the semantic space in LLM and the recommendation space in the multi-scenario backbone model. On the other hand, to tackle the efficiency of LLM (see Section \ref{method:disc}), a frozen LLM is used without tuning and high level of user- \& scenario knowledge is reasoned about through prompt design.

The main contributions of this paper are summarized as follows:
\begin{itemize}[leftmargin=*]
\item We propose an effective efficient interpretable paradigm \name for multi-scenario recommendation enhancement by exploiting LLM. To the best of our knowledge, it is the first practical solution to reasoning about multi-scenario knowledge
through LLM and bring information gain in recommender systems.

\item The scenario commonality and heterogeneity, and users' cross-scenario preferences are explicitly captured by LLM through the designed scenario- and user-level prompt. Furthermore, they are leveraged by the hierarchical meta networks which generate meta layers to explicitly enhance the scenario-aware and personalized recommendation capability of multi-scenario backbones. 

\item Extensive experiments on three real-world datasets show that \name is an effective paradigm compared with state-of-the-art multi-scenario enhancement methods and compatible with various MSR backbone models. Besides, it is efficient to deploy on industrial recommender systems and enables real-time recommendation \textbf{without fine-tuning LLM}. 
\end{itemize}
\section{Preliminary} \label{prel}
This section first introduces the problem formulation of multi-scenario CTR prediction and the typical architecture of multi-scenario backbone models. Then, large language model is illustrated.

\subsection{Multi-Scenario CTR Prediction}
The Click-Trough Rate (CTR) prediction task is essential in recommendation which is a binary classification problem. Specifically, in a multi-scenario recommender system, the historical interaction sample $(d,\boldsymbol{x})$ is taken as input to predict the ground truth label $y$ where $y=1$ and $y=0$ denote click and unclick, respectively. $d$ represents the domain indicator $d \in \{1,2,\dots,D\}$ indicating the origin of sample. $\boldsymbol{x}$ denotes the feature fields including user and item attributes. For simplicity, suppose there are $M$ categorical features, then $\boldsymbol{x}=[\boldsymbol{x}_1, \ldots, \boldsymbol{x}_m, \ldots, \boldsymbol{x}_M]$ where $\boldsymbol{x}_m$ indicates the one-hot encoding of the $m$-th feature. Next, $\boldsymbol{x}$ is mapped into a dense vector $\boldsymbol{e}=[\boldsymbol{e}_1 \Vert \ldots \Vert \boldsymbol{e}_m \Vert \ldots \Vert \boldsymbol{e}_{M}]$ through an embedding layer where $\Vert$ denotes the concatenation operation. For the $m$-th feature, we calculate $\boldsymbol{e}_{m}$ through a look-up operation $\boldsymbol{e}_{m}=\boldsymbol{E}_m \cdot \boldsymbol{x}_m$ in which $\boldsymbol{E}_m \in \mathbb{R}^{u_m \times Dim}$ denotes the embedding table, $u_m$ denotes the value counts of feature, and $Dim$ is the embedding size. Finally, the prediction result $\hat{y}$ is obtained via $\hat{y}=f_d(\boldsymbol{e})$ where $f_d$ represents the recommendation model applied to the $d$-th scenario. The loss function is binary cross entropy loss or the so-called Logloss \cite{berkson1944application}:
\begin{equation} \label{eq1}
\setlength{\abovedisplayskip}{6pt}
\setlength{\belowdisplayskip}{6pt}
\underset{\Theta}{\min} \mathcal{L}=-\frac{1}{B} \sum_{i=1}^B y_i \log \left(\hat{y}_i\right)+\left(1-y_i\right) \log \left(1-\hat{y}_i\right),
\end{equation}
where $\Theta$ is the set of parameters learned, $B$ represents the number of batch samples, and $y_i$ and $\hat{y}_i$ denote the true label and prediction of the $i$-th sample, respectively.

\begin{figure}[t]
\centering
\setlength\abovecaptionskip{0\baselineskip}
\setlength\belowcaptionskip{0.5\baselineskip}
\includegraphics[scale=0.48, trim=0 0 24 0,clip]{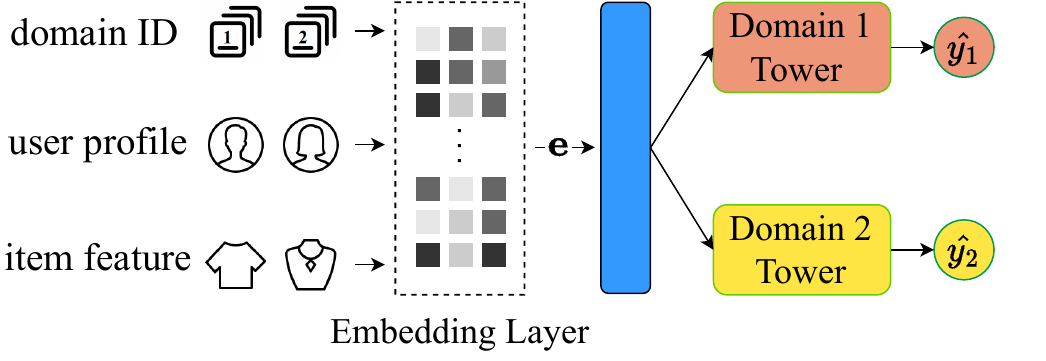}
\caption{Demonstration of multi-scenario recommender system. The blue part represents shared parameters while red and yellow denote scenario-specific parameters.}
\label{model}
\end{figure}

As depicted in Figure \ref{model}, most multi-scenario recommendation models model the commonality and distinction among scenarios through explicitly maintaining domain-shared parameters $\Theta_s$ like the experts in MMoE \cite{ma2018modeling} and domain-specific parameters $\Theta_u$ like towers and gates in PLE \cite{tang2020progressive}.  


\section{Proposed Method} \label{meth}

\begin{figure*}[t]
\centering
\setlength\abovecaptionskip{0.2\baselineskip}
\setlength\belowcaptionskip{0.2\baselineskip}
\includegraphics[scale=0.26, trim=0 0 45 0,clip]{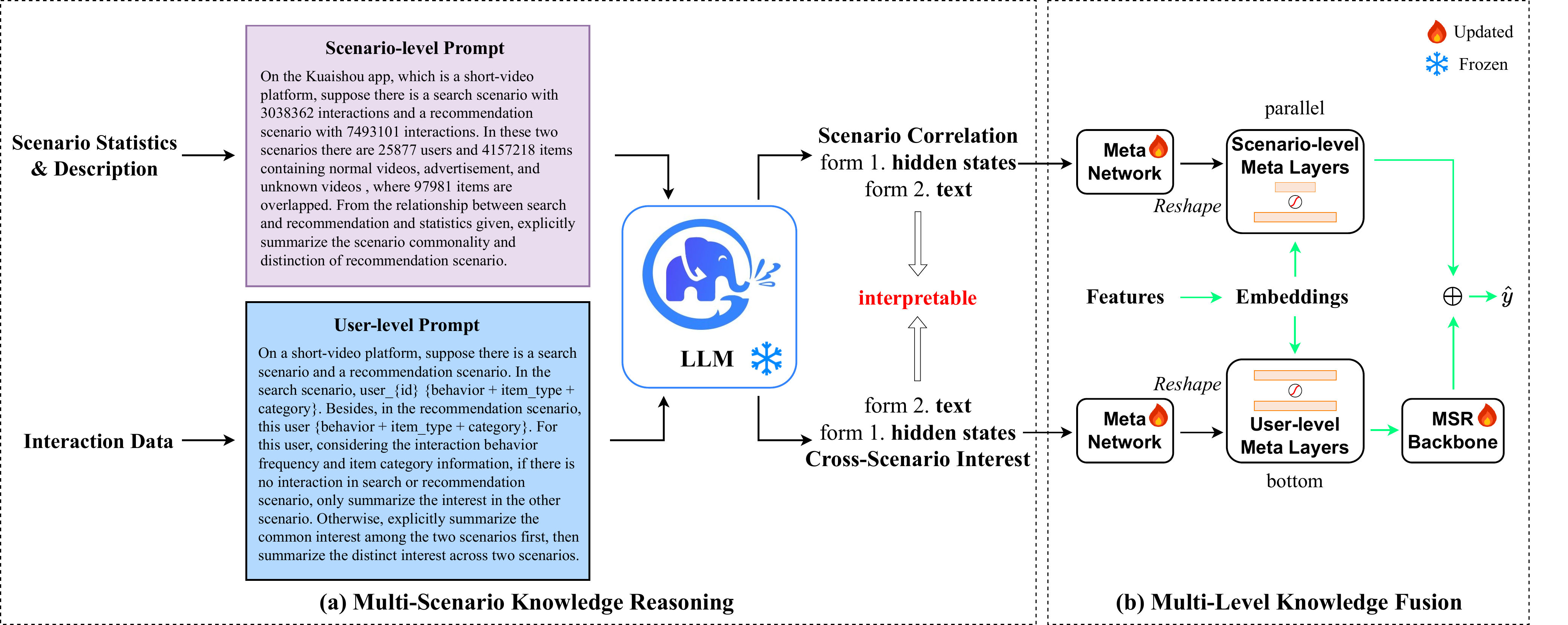}
\caption{Overall framework of \name. Different colors of solid arrows indicate different data flow. In multi-scenario knowledge reasoning step, an LLM captures scenario correlation and cross-scenario preferences from scenario- and user-level prompt without fine-tuning. In multi-level knowledge fusion step, two hierarchical meta networks, which adaptively generate scenario- and user-level meta layers in parallel and at bottom, are end-to-end trained with the multi-scenario backbone model.}
\label{framework}
\vspace{-2mm}
\end{figure*}

We first illustrate the overall framework of \name. Next, it is further detailed in the multi-scenario knowledge reasoning and multi-level knowledge fusion step. Finally, we discuss the idea and characteristics of \name compared with the existing multi-scenario and LLM-based recommendation methods, and illustrate the necessity of LLMs versus Pre-trained Language Models (PLMs).

\subsection{Overall Framework}

To tackle the limitations of the existing MSR methods, i.e., insufficient scenario knowledge incorporation and ignoring personalized cross-scenario interest, we propose a paradigm \name for multi-scenario enhancement which is compatible with different multi-scenario backbone models.
Specifically, it is equipped with an LLM with frozen parameters and two learnable meta networks.
Meanwhile, the general procedure of \name is summarized in Algorithm \ref{alg}.
It is comprised of two steps:
\textbf{(i)} multi-scenario knowledge reasoning (from Line 1 to 10) including scenario correlation and cross-scenario interests inference, and
\textbf{(ii)} multi-level knowledge fusion (from Line 11 to 15) to adaptively fuse with backbone.

\begin{algorithm}[t]

\caption{\name paradigm taking multi-scenario CTR prediction as an example} \label{alg}

\KwIn{Domain ID $d \in \{1,2,\dots,D\}$; set of user $u \in \{1,2,\dots,U\}$; features $\boldsymbol{x}$; true label $y$; a pre-trained large language model (LLM)}
\KwOut{A well trained model for all scenarios}

\nonl Step 1: Multi-scenario Knowledge Reasoning\

\For{$d \in \{1,2,\dots,D\}$}{
 Generate scenario-level prompt\;
 Feed into LLM\;
 Obtain the last hidden state of scenario-level output\;
}

\For{$u \in \{1,2,\dots,U\}$}{
 Generate user-level prompt\;
 Feed into LLM\;
 Obtain the last hidden state of user-level output\;
}

\nonl Step 2: Multi-level Knowledge Fusion\

\While{not converge}{
 Sample a mini-batch data from all scenarios\;

 Calculate the corresponding scenario- and user-level meta layers via Equation \eqref{eq4_1} and \eqref{eq4_2}\;
 Calculate the prediction result via Equation \eqref{eq7}\;
 Calculate the loss via Equation \eqref{eq1}\; 
 Take the gradient and update parameters\;
}
\end{algorithm}

The illustration of \name is depicted in Figure \ref{framework}. Specifically, the scenario-level prompt is generated from the scenario statistics, semantic descriptions, and expert knowledge while the user-level prompt is constructed using historical interactions.
Afterward, they are fed into LLM to generate  scenario correlation and personalized cross-scenario interest respectively. 
Notably, only \textbf{the last hidden state} of LLM output (i.e., the hidden state of the last token generated) is chosen as representative and stored which is a high-dimensional dense vector, i.e., in 4096 dimensions taking ChatGLM2-6B \cite{du2022glm} as an example. It is common because the generation of the last token is based on all previous ones and contains overall information.
Finally, the hierarchical meta networks produce meta layers in scenario- and user-level, and they are trained with the MSR backbone model to enhance its all-scenario performance.
Overall, the final enhanced model can be obtained in a simple yet efficient end-to-end training manner by optimizing Equation \eqref{eq1}.

\vspace{-0.25em}
\subsection{Multi-Scenario Knowledge Reasoning} \label{method:pd}
On the one hand, it is difficult for the mainstream MSR models to grasp the complex scenarios correlations because in practice the domain indicator distinguishing data origin is usually the only scenario-related knowledge \cite{sheng2021one,chang2023pepnet,gao2023scenario}, thus their scenario representation is only learned from the implicit collaborative signals.
Therefore, we design a scenario-level prompt to explicitly capture the commonality and distinction of scenarios from their statistics and semantic descriptions.
For example, as shown in the purple box in Figure \ref{framework}, the statistics like the number of interactions, users, items, and overlaps on KuaiSAR-small together with the background semantic information are given to uncover both similarity and heterogeneity of the recommendation and search scenario.

On the other hand, existing MSR models tend to neglect modeling personalized interest and it is challenging to capture user's cross-scenario preferences in a disentangled manner. To this end, we design a user-level prompt to explicitly explore personalized common and distinct interests among all the scenarios from the abundant historical interaction.
Intuitively, positive interaction behavior tend to better reflect user's interest than negative behavior.
Meanwhile, the proportion of negative samples is usually large and they contain much noise information due to the biases and uncertainty in data collection and user interaction.
Consequently, the positive interaction data (i.e., samples with positive interaction behavior such as click, forward, and like) are filtered and leveraged in the user-level prompt.
As shown in the blue box in Figure \ref{framework}(a), LLM helps explicitly capturing user's preferences in the search and recommendation scenario considering both the interaction frequency and the semantic relationships, e.g., in the category and title of item. The specific prompt template are provided in Appendix \ref{pt}.

After obtaining the scenario- and user-level prompt, they are passed into LLM to reason about the multi-scenario knowledge. Moreover, we conduct a case study in Section \ref{cs} to investigate what multi-scenario knowledge can LLM obtain and how it contributes to enhancing the MSR backbone model.

\vspace{-0.25em}
\subsection{Multi-Level Knowledge Fusion} \label{method:mn}
Existing methods either choose to directly leverage the augmented information from LLM as additional input features in the RS \cite{xi2023towards},  adopt a linear projection or a fully connected network to conduct dimension transformation and align instance representation \cite{li2023ctrl}, or fine-tune the LLM \cite{zheng2023adapting}.
However, we argue that these are sub-optimal solutions \cite{su2023stem}.
First, since the LLM-augmented information is treated as a common feature by recommender system, it is unable to play a significant role as expected because the number of feature fields is usually large \cite{zhao2021autodim}.
Second, simply aligning the dimension would result in loss of information \cite{zebari2020comprehensive}. 
Third, tuning LLM brings high cost \cite{chen2023extending,chen2023longlora} and it requires consistent update.

Therefore, after obtaining the informative multi-scenario knowledge which is in the form of high-dimensional hidden vectors from LLM, our proposed \name adopts a new and efficient approach leveraging hierarchical meta networks to bring information gain.
To be specific, the meta networks dynamically generate meta layers to adaptively fuse the scenario- and user-level knowledge with the collaborative information in the MSR backbone model. 
Intuitively, on the one hand, the scenario-level knowledge usually plays a more general role like the domain distinction captured by auxiliary network in STAR \cite{sheng2021one}.
On the other hand, the user-level information tends to be more personalized and significant \cite{chang2023pepnet} because it breaks the boundary of scenarios and contains a complete user profile across all scenarios.
Consequently, we propose a hierarchical \textbf{`bottom + parallel'} structure, in which the user-level 
meta layers are placed at the bottom while the scenario-level meta layers are implemented in parallel to the MSR backbone, as shown in Figure \ref{framework}(b). 
Notably, They target at improving scenario-aware recommendation and
cross-scenario personalization capability respectively, thus contributing to knowledge fusion and enhancement in different levels and granularity.
We will empirically investigate this architecture of \name through experiments in Section \ref{abc} where we also observe that user-level meta layers at bottom and scenario-level meta layers in parallel is the optimal combination.

Specifically, denote the last hidden state of LLM output as $\boldsymbol{h}_{LLM}$, the meta work takes it as input and outputs $\boldsymbol{h}_{mw}\in \mathbb{R}^{Dim_{mw}}$ and $\boldsymbol{h}_{mb} \in \mathbb{R}^{Dim_{mb}}$, then a reshape operation is employed to obtain $K$ meta layers in total:
\begin{equation}
\setlength{\abovedisplayskip}{3pt}
\setlength{\belowdisplayskip}{1pt}
\boldsymbol{h}_{mw}, \boldsymbol{h}_{mb}=\operatorname{Meta~Network}(\boldsymbol{h}_{LLM}),
\end{equation}
\begin{equation} \label{eq4_1}
\setlength{\abovedisplayskip}{3pt}
\setlength{\belowdisplayskip}{1pt}
\boldsymbol{W}_{l}^{(i)}=Reshape(\boldsymbol{h}_{mw})
\end{equation}
\begin{equation} \label{eq4_2}
\setlength{\abovedisplayskip}{3pt}
\setlength{\belowdisplayskip}{1pt}
\boldsymbol{b}_{l}^{(i)}=Reshape(\boldsymbol{h}_{mb}), i \in \{1,2,\ldots,K\}
\end{equation}
where $\boldsymbol{W}_{l}^{(i)}$ and $\boldsymbol{b}_{l}^{(i)}$ indicate the weight and bias matrix of the $i$-th meta layer. Notably, $Dim_{mw}$ and $Dim_{mb}$ equals to the sum of dimensions of all $\boldsymbol{W}_{l}^{(i)}$ and $\boldsymbol{b}_{l}^{(i)}$, respectively. For example, $\boldsymbol{h}_{mw}$ with the dimension of 8256 can be reshaped into a weight matrix with dimension 128$\times$64 plus a weight matrix with dimension 64$\times$1 (bias matrix is similar). Then each meta layer is followed by an activation function $\sigma$ like ReLU:
\begin{equation}
\setlength{\abovedisplayskip}{6pt}
\setlength{\belowdisplayskip}{6pt}
\boldsymbol{h}^{(i)}=\sigma(\boldsymbol{W}_{l}^{(i)}\boldsymbol{h}^{(i-1)} + \boldsymbol{b}_{l}^{(i)}), i \in \{1,2,\ldots,K\}
\end{equation}
where $\boldsymbol{h}^{(i)}$ is the output of the $i$-th meta layer.
Afterward, the feature embedding $\boldsymbol{e}$, which is exactly $\boldsymbol{h}^{(0)}$, is transferred into the obtained meta layers. Finally, the prediction result is calculated by combining the output of the parallel scenario-level meta layers and the result from MSR backbone  
in an adaptive manner:
\begin{equation}
\setlength{\abovedisplayskip}{3pt}
\setlength{\belowdisplayskip}{1pt}
\boldsymbol{h}=\operatorname{MSR}(\boldsymbol{h}_u^{(K)}),
\end{equation}
\begin{equation} \label{eq7}
\setlength{\abovedisplayskip}{1pt}
\setlength{\belowdisplayskip}{3pt}
\hat{y}=\sigma^{\prime}(\alpha \cdot \boldsymbol{h}_s^{(K)}+(1-\alpha) \cdot \boldsymbol{h}),
\end{equation}

\noindent 
where $\boldsymbol{h}$ is the output hidden vector of the MSR backbone model, $\boldsymbol{h}_u^{(K)}$ is the output of the $K$-th user-level meta layer, $\boldsymbol{h}_s^{(K)}$ is the output of the $K$-th scenario-level meta layer, $\sigma^{\prime}$ is the activation function for the final task (e.g., sigmoid function for CTR prediction task), and $\alpha$ is a learnable parameter.

\subsection{Discussion} \label{method:disc}

In the following part, we first illustrate the idea of the proposed \name paradigm, then compare it with the mainstream multi-scenario RSs and the LLM-based recommendation methods, and finally demonstrate the necessity of LLM here compared with PLM.

Basically, rather than designing a specific multi-scenario recommendation model architecture, our goal is to propose a general multi-scenario enhancement paradigm.
Specifically, considering the model architecture and optimization, \name does not impose any restrictions on the multi-scenario backbone models, thus being model-agnostic.
Meanwhile, for LLM the most two significant questions are: (i) \textbf{WHAT} information can be obtained from LLM and (ii) \textbf{HOW} to use it to bring information gain, which correspond to the multi-scenario knowledge reasoning and multi-level knowledge fusion steps proposed in our \name.

\textbf{\name v.s. multi-scenario RS.}
Conventional multi-scenario RSs treat the domain indicator either as a distinct feature \cite{sheng2021one} or as the only scenario knowledge to generate scenario-specific module \cite{gao2023scenario}, and adopt different parameter sharing patterns to model scenario correlations where only the collaborative signal is learned. 
By contrast, \name generates LLM-enhanced scenario- and user-level representation from the semantic space, then it dynamically generates meta layers to effectively and efficiently combine with the multi-scenario RS based on these representations.
Consequently, the LLM-augmented information is flexibly fused with the collaborative signal in conventional RS, thus achieving enhancement.

\textbf{\name v.s. LLM-based RS.} 
Existing LLM-based recommender systems can be categorized into two groups, i.e., using LLM as the knowledge enhancer \cite{li2023ctrl,wang2023alt,lin2023clickprompt,gong2023unified} and directly as the recommender \cite{geng2022recommendation,cui2022m6,liu2023chatgpt}.
However, the former models possess low efficiency since they require aligning the information from LLM and RS in the instance level.
Meanwhile, the latter methods lack scenario knowledge, thus achieving unsatisfying recommendation performance and suffering from high inference latency of LLM.

By contrast, \name can efficiently and effectively tackle these problems.
First, thanks to the zero-shot capability \cite{wang2022language}, an LLM with fixed parameters is leveraged free from tuning in \cite{yin2023heterogeneous,li2023ctrl,wang2023alt}.
Meanwhile, the way of obtaining information from LLM is shifted from the low \textbf{`instance-level'} (e.g., in \cite{li2023ctrl,wang2023alt,zheng2023adapting}) to the high \textbf{`user- \& scenario-level'}, thus improving efficiency.
Besides, through hierarchical meta networks, the multi-level scenario knowledge is adaptively injected and lead to enhancement.
Moreover, apart from distinct scenario-specific and user-specific features, it also enables to incorporate expert scenario knowledge through the prompt design, which is more universal and advanced than simply extracting invariant features and aligning the heterogeneous ones in \cite{gong2023unified}.


\textbf{Necessity of LLMs v.s. PLMs.} 
Unlike PLMs like BERT trained with masked language modeling task, most popular LLMs like ChatGLM2 \cite{du2022glm} 
are trained with next token prediction objective. 
Therefore, rather than simply deriving a representation of existing (old) information by adopting PLMs as \textbf{encoder} (like BERT in CTRL \cite{li2023ctrl}), the key of leveraging LLM as \textbf{reasoner} in \name is to further infer useful (new) knowledge like analyzing user's cross-scenario preference. This knowledge (in the form of hidden vectors) could potentially help enhancing conventional recommender systems. Besides, a case study is conducted in Section \ref{cs} and the knowledge (in the form of text) from ChatGLM2-6B is shown in Figure \ref{fig:RQ5}, which helps improving interpretability.

By contrast, we also adopt GPT-2 (a typical PLM) to reason about user\_1's interest given the same input prompt as in ChatGLM2-6B, and it generates the following nonsense text:

\begin{verbatim}
"The main strategy of this study was to combine two 
scenarios into a two-way analysis. The primary strategies 
are presented in Table I. If the main scenario is presented 
as a list of stories, then..."
\end{verbatim}
In that case, such information could not bring information gain and improve the recommendation performance.


\section{Experiments} \label{expe}

Extensive experiments are conducted on three public datasets to verify the effectiveness and efficiency of our proposed LLM4MSR paradigm, and  the following five questions are answered:
\begin{itemize}[leftmargin=*]
\vspace{1mm}
\item \textbf{RQ1:} Is LLM4MSR an effective paradigm and compatible with different multi-scenario backbone models?
\item \textbf{RQ2:} What is the effect of the scenario-level and user-level prompt? What is their optimal architecture correspondingly?
\item \textbf{RQ3:} What is the impact of interaction threshold and the number of neural network layers generated by the meta network?
\item \textbf{RQ4:} What is the efficiency of LLM4MSR compared with the original multi-scenario backbone models?
\item \textbf{RQ5:} How does LLM help improving the multi-scenario backbone models through our proposed LLM4MSR?
\end{itemize}

\subsection{Experimental Settings}
\subsubsection{\textbf{Datasets}} Our experiments are conducted on three datasets: KuaiSAR-small, KuaiSAR \cite{Sun2023KuaiSAR}, and Amazon \cite{ni2019justifying}. To be specific, there is a search and a recommendation scenario on KuaiSAR-small and KuaiSAR while there are three correlated recommendation scenarios 
selected on Amazon. Furthermore, their statistics and detailed descriptions are
summarized in Table \ref{tab:stat}. To be specific, we divide the original data into training, validation, and test set with the ratio of 8:1:1 by chronological order. Besides, the sparsity metric denotes the proportion of unclicked samples with label $y=0$.

\begin{itemize}[leftmargin=*]
\vspace{1mm}
\item \textbf{KuaiSAR-small \& KuaiSAR\footnote{https://kuaisar.github.io/}} datasets are both crawled from the short-video platform Kuaishou. The only difference is KuaiSAR-small is a smaller version of data in 9 days while KuaiSAR is in 19 days. Specifically, all users are overlapped in the recommendation and search scenario. Meanwhile, there are 97981 and 181849 items overlapped on KuaiSAR-small and KuaiSAR, respectively.

\item \textbf{Amazon\footnote{https://cseweb.ucsd.edu/{\textasciitilde}jmcauley/datasets/amazon\_v2/}} is collected from Amazon. Following \cite{cui2020herograph}, we select three connected
scenarios: All Beauty, Clothing Shoes and Jewelry, and Luxury Beauty and we also filter the overlapped users among these scenarios without overlapped items.
The goal is to predict whether a user would give a rating (from 1 to 5) higher than 3 to an item.

\end{itemize}

\begin{table}[t] \small
\centering
\setlength\abovecaptionskip{0.1\baselineskip}
\setlength\belowcaptionskip{-0.3\baselineskip}
\caption{The statistics of KuaiSAR-small, KuaiSAR, and Amazon, where Rec denotes recommendation.}
    \label{tab:stat}
    \resizebox{0.475\textwidth}{!}{
    \setlength{\tabcolsep}{1mm}{
    \begin{tabular}{cccccccccc}
    \toprule
        Dataset & \multicolumn{2}{c}{KuaiSAR-small} & \multicolumn{2}{c}{KuaiSAR} & \multicolumn{3}{c}{Amazon}  \\
    \midrule    
        Scenarios & Rec & Search & Rec & Search & Rec$\#1$ & Rec$\#2$ & Rec$\#3$ \\
        Users & 25,877 & 25,877 & 25,877 & 25,877 & 24,752 & 24,752 & 24,752  \\
        Items & 2,281,034 & 1,974,165 & 4,046,363 & 2,974,596 & 8,788 & 193,304 & 6,980 \\
        Interactions & 7,493,101 & 3,038,362 & 14,605,712 & 4,828,690 & 33,929 & 370,840 & 56,356 \\
        Sparsity & 50.49\% & 87.87\% & 50.40\% & 87.88\% & 21.50\% & 19.61\% & 20.67\% \\
    \bottomrule
    \end{tabular}}
    \vspace{3mm}
    }
\end{table}

\subsubsection{\textbf{Evaluation Metrics}} Area Under the ROC curve (AUC) is adopted to conduct evaluation on the test set, where a higher AUC value at 0.001 level usually denotes significant enhancement \cite{cheng2016wide,wang2023plate}.

\subsubsection{\textbf{Backbone Models}} \label{backbone}

We implement \name on the following representative multi-scenario backbone models commonly deployed in industrial recommender system:
\begin{itemize}[leftmargin=*]
\vspace{1mm}
\item \textbf{STAR} \cite{sheng2021one} proposes a star topology architecture with a shared centered network multiplied by many domain-specific networks.
\item \textbf{OMoE} \cite{jacobs1991adaptive} leverages a group of expert networks and a gating network learned to ensemble the results from these experts.
\item \textbf{MMoE} \cite{ma2018modeling} explicitly learns a gating network for each task to assemble task-specific information compared with OMoE.
\item \textbf{PLE} \cite{tang2020progressive} proposes Customized Gate Control (CGC) which separates shared and specific expert networks and employs multi-level extraction network with a progressive routing mechanism.
\item \textbf{AITM} \cite{xi2021modeling} adaptively learns to transfer information for different tasks with sequential dependence.
\item \textbf{Shared Bottom} is a multi-task model where the embedding layer is usually shared on which task-specific networks are built.
\end{itemize}

\begin{table*}[t]
\centering
\setlength\abovecaptionskip{0.2\baselineskip}
\setlength\belowcaptionskip{0.2\baselineskip}
\caption{Overall performance comparison of AUC on KuaiSAR-small, KuaiSAR, and Amazon dataset. The suffix `DN', `EP', and `ours' denote Dynamic Network, EPNet, and \name. Boldface denotes the highest value while underline indicates the second best result. $\star$ represents statistical significance with $p$-value $< 0.05$ in $t$-test compared with the backbone model.}
\label{tab:overall}

\resizebox{0.7\textwidth}{!}{
\begin{tabular}{@{}cccccccc@{}}
\toprule
\multirow{2}{*}{{Backbones}} &  \multicolumn{2}{c}{KuaiSAR-small} & \multicolumn{2}{c}{KuaiSAR} & \multicolumn{3}{c}{Amazon} \\ \cmidrule(l){2-8}
 & Rec & Search & Rec & Search & Rec$\#1$ & Rec$\#2$ & Rec$\#3$ \\ \midrule
STAR & 0.7225 & 0.6089 & 0.7387 & 0.6268 & 0.6156 & 0.6320 & 0.6225 \\
STAR\_DN & {\ul 0.7241} & {\ul 0.6116} & {\ul 0.7404} & {\ul 0.6270} & {\ul 0.6306} & {\ul 0.6350} & 0.6355 \\
STAR\_EP & 0.7230 & 0.6082 & 0.7388 & 0.6266 & 0.6211 & 0.6302 & {\ul 0.6378} \\
STAR\_ours & \bm{$0.7276^{\star}$} & \bm{$0.6181^{\star}$} & \bm{$0.7408^{\star}$} & \bm{$0.6332^{\star}$} & \bm{$0.8720^{\star}$} & \bm{$0.6364^{\star}$} & \bm{$0.7543^{\star}$} \\
\midrule
OMoE & 0.7241 & 0.6163 & 0.7394 & 0.6310 & 0.5995 & 0.6043 & 0.6252 \\
OMoE\_DN & {\ul 0.7249} & {\ul 0.6183} & {\ul 0.7402} & 0.6319 & 0.6027 & 0.6171 & 0.6289 \\
OMoE\_EP & 0.7246 & 0.6179 & 0.7392 & {\ul 0.6330} & {\ul 0.6143} & {\ul 0.6176} & {\ul 0.6312} \\
OMoE\_ours & \bm{$0.7265^{\star}$} & \bm{$0.6186^{\star}$} & \bm{$0.7413^{\star}$} & \bm{$0.6332^{\star}$} & \bm{$0.8182^{\star}$} & \bm{$0.6180^{\star}$} & \bm{$0.6823^{\star}$} \\
\midrule
MMoE & 0.7235 & 0.6150 & 0.7392 & 0.6313 & 0.5907 & 0.5999 & 0.6032 \\
MMoE\_DN & {\ul 0.7246} & 0.6164 & 0.7394 & {\ul 0.6330} & {\ul 0.6310} & 0.6201 & {\ul 0.6371} \\
MMoE\_EP & 0.7245 & \bm{$0.6178$} & {\ul 0.7397} & 0.6326 & 0.6218 & {\ul 0.6221} & 0.6315 \\
MMoE\_ours & \bm{$0.7264^{\star}$} & {\ul $0.6166^{\bm{\star}}$} & \bm{$0.7410^{\star}$} & \bm{$0.6341^{\star}$} & \bm{$0.7860^{\star}$} & \bm{$0.6222^{\star}$} & \bm{$0.6854^{\star}$} \\
\midrule
PLE & 0.7249 & 0.6149 & 0.7396 & 0.6313 & 0.6059 & 0.6127 & 0.5995  \\
PLE\_DN & {\ul 0.7258} & {\ul 0.6165} & 0.7400 & 0.6333 & 0.6066 & 0.6195 & 0.6162  \\
PLE\_EP & 0.7252 & \bm{$0.6184$} & {\ul 0.7402} & {\ul 0.6339} & {\ul 0.6173} & {\ul 0.6229} & {\ul 0.6263}  \\
PLE\_ours & \bm{$0.7269^{\star}$} & \bm{$0.6184^{\star}$} & \bm{$0.7408^{\star}$} & \bm{$0.6343^{\star}$} & \bm{$0.8265^{\star}$} & \bm{$0.6250^{\star}$} & \bm{$0.7074^{\star}$}  \\
\midrule
AITM & 0.7236 & 0.6154 & 0.7398 & 0.6329 & 0.6039 & 0.6126 & 0.6046  \\
AITM\_DN & {\ul 0.7243} & 0.6172 & {\ul 0.7398} & 0.6318 & 0.6057 & 0.6151 & 0.6166  \\
AITM\_EP & 0.7237 & {\ul 0.6177} & 0.7389 & {\ul 0.6337} & {\ul 0.6064} & {\ul 0.6163} & {\ul 0.6238}  \\
AITM\_ours & \bm{$0.7273^{\star}$} & \bm{$0.6178^{\star}$} & \bm{$0.7407^{\star}$} & \bm{$0.6349^{\star}$} & \bm{$0.8196^{\star}$} & \bm{$0.6178^{\star}$} & \bm{$0.7089^{\star}$} \\
\midrule
Shared Bottom & 0.7228 & 0.6169 & 0.7389 & 0.6321 & 0.6073 & 0.6077 & 0.6268  \\
Shared Bottom\_DN & {\ul 0.7250} & {\ul 0.6176} & {\ul 0.7396} & 0.6323 & 0.6081 & {\ul 0.6253} & {\ul 0.6294}  \\
Shared Bottom\_EP & 0.7243 & \bm{$0.6182$} & 0.7392 & {\ul 0.6331} & {\ul 0.6268} & 0.6174 & 0.6233  \\
Shared Bottom\_ours & \bm{$0.7269^{\star}$} & \bm{$0.6182^{\star}$} & \bm{$0.7400^{\star}$} & \bm{$0.6341^{\star}$} & \bm{$0.8323^{\star}$} & \bm{$0.6262^{\star}$} & \bm{$0.7182^{\star}$} \\

\bottomrule
\end{tabular}}
\end{table*}

\subsubsection{\textbf{Baselines}} \label{baseline}

There are few works aimed at explicitly enhancing multi-scenario recommendation. Besides, previous LLM-based models are not applicable in the same setting (refer to Section \ref{sec:llm4rec}), e.g., LLM as recommender fails to meet real-time inference requirement. Consequently, we compare 
\name with 
two 
typical paradigm as baseline methods:
\begin{itemize}[leftmargin=*]
\vspace{1mm}
\item \textbf{Dynamic Network}. It targets at adaptively generating model parameters in a fine-grained manner \cite{yan2022apg}. In multi-scenario modeling \cite{zhang2022leaving}, the scenario knowledge is employed to generate scenario-specific attention and tower module.
\item \textbf{EPNet}. It is a gate structure proposed in PEPNet \cite{chang2023pepnet}. Specifically, it takes domain-side information as input and produce scenario-specific gates to transform embedding. 
\end{itemize} 

\subsubsection{\textbf{Implementation Details}} 
The implementation details of the experiments conducted are summarized in \textbf{Appendix \ref{detail}} 
and the code is available to ease reproduction\textsuperscript{\ref{foot1}}. Specifically, for the open-source LLM we choose ChatGLM2-6B\footnote{\label{foot2}\url{https://huggingface.co/THUDM/chatglm2-6b}} \cite{du2022glm} as an example. 


\subsection{Overall Performance (RQ1)}

To answer \textbf{RQ1}, we verify \name's compatibility with various multi-scenario backbone models introduced in Section \ref{backbone}. Meanwhile, \name is compared with two methods for multi-scenario recommendation enhancement as the baseline methods detailed in Section \ref{baseline}. The overall performance of AUC on KuaiSAR-small, KuaiSAR, and Amazon dataset is shown in Table \ref{tab:overall}.

First, Dynamic Network and EPNet are able to simultaneously improve the recommendation accuracy on different scenarios in most cases, while there are exceptions like `AITM\_DN' on KuaiSAR and `STAR\_EP' on all datasets. By contrast, the proposed \name significantly enhances the recommendation performance in all scenarios on the three datasets especially in Rec\#1 and Rec\#3 on Amazon. It also consistently outperforms the baseline methods except in the search scenario on KuaiSAR-small based on MMoE. The reason for its superiority is that,
on the one hand, both the multi-scenario backbone models and baseline methods incorporate insufficient scenario knowledge, i.e., domain id only, and lack personalized  modeling.
On the other hand, \name first utilizes LLM to explore domain correlation and infer personalized cross-scenario interest, then integrates meta network to explicitly enhance multi-scenario modeling capability.
Besides, we observe that compared with the significant enhancement on scenario Rec\#1 and Rec\#3, \name results in relatively subtle improvement (around 0.7\%) on scenario Rec\#2 on Amazon. The reason is that Rec\#1 and Rec\#3 are relatively sparse where each user has 1.4 and 2.3 interactions on average as shown in Table \ref{tab:stat}. Therefore, less information of user preference on these two scenarios could be captured by LLM to enhance Rec\#2.

Overall, \name achieves an increase of 1.5\%, 1\%, and 40\% in AUC on three datasets. 
To summarize, we can draw the conclusion that \name serves as an effective and powerful enhancement paradigm, which possesses great compatibility and deployability on the mainstream multi-scenario recommendation backbone models.

\subsection{Ablation \& Component Study (RQ2)} \label{abc}

\begin{figure}[t]
\setlength\abovecaptionskip{-0.6\baselineskip}
\setlength\belowcaptionskip{0.3\baselineskip}
	\centering
	\begin{minipage}{0.345\linewidth}
		\centering
        \begin{subfigure}{1\linewidth}
		\includegraphics[width=0.995\linewidth, trim=16 12 6 6,clip]{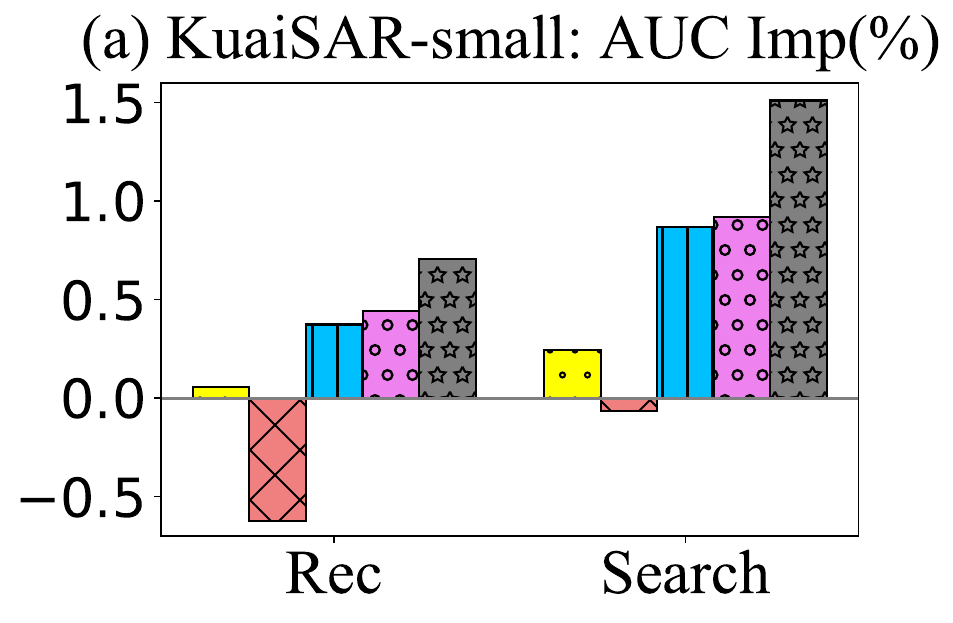}
		\label{fig:RQ2-1}
        \end{subfigure}
	\end{minipage}
	\begin{minipage}{0.317\linewidth}
		\centering
        \begin{subfigure}{1\linewidth}
		\includegraphics[width=0.995\linewidth, trim=16 12 6 6,clip]{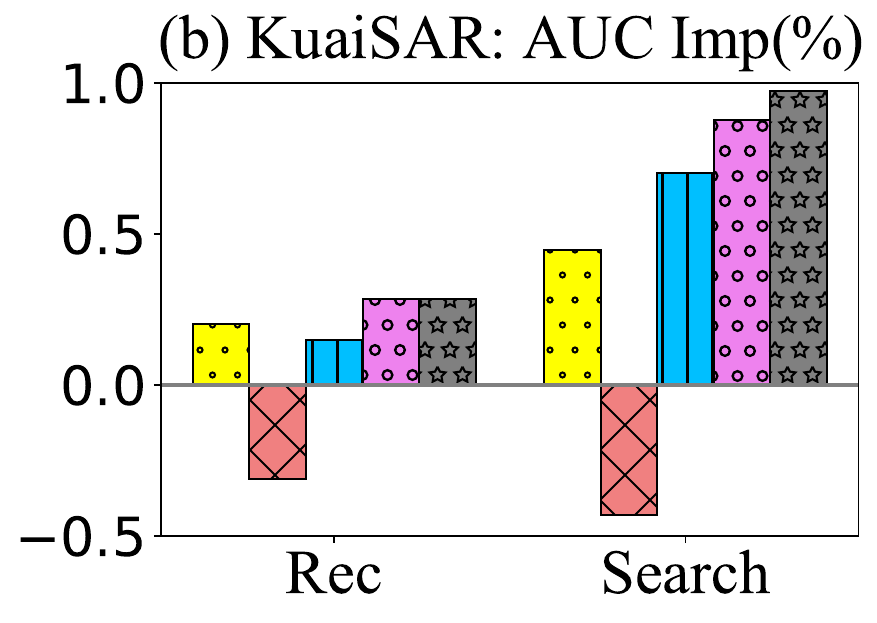}
		\label{fig:RQ2-2}
        \end{subfigure}
	\end{minipage}
        \begin{minipage}{0.317\linewidth}
		\centering
        \begin{subfigure}{1\linewidth}
		\includegraphics[width=0.995\linewidth]{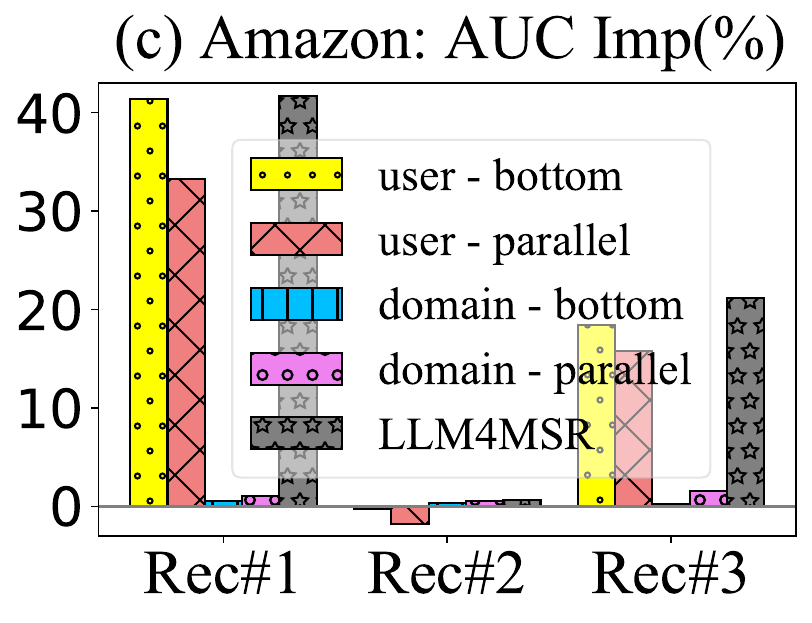}
		\label{fig:RQ2-3}
        \end{subfigure}
	\end{minipage}
	\caption{Ablation and component study on KuaiSAR-small, KuaiSAR, and Amazon dataset. The prefix `user' and `domain' denotes user-level and domain-level prompt while the suffix `bottom' and `parallel' indicates the architecture applied. We adopt the ratio of improvement in AUC value compared with the original STAR model as the backbone.} 
        \vspace{-6.5mm}
	\label{fig:RQ2}
\end{figure}

To answer \textbf{RQ2}, we conduct ablation study and component analysis based on STAR considering the following variants. 
\begin{itemize}[leftmargin=*]
\item user-bottom / user-parallel: It only adopts user-level meta layers at bottom / in parallel \textbf{without} scenario knowledge as input.
\item domain-bottom / domain-parallel: It only adopts scenario-level meta layers at bottom / in parallel \textbf{without} user knowledge.
\item \name: Full model with both user-bottom and domain-parallel.
\end{itemize}
The results are depicted in Figure \ref{fig:RQ2}. 
On the one hand, the full model \name achieves the best result on all scenarios on all the three datasets. The reason is that it makes full use of both the informative multi-scenario knowledge and the collaborative signal learned from the LLM and MSR backbone model, respectively. Meanwhile, from `user - bottom' and `domain - parallel' it is seen that both user- and domain-level prompt contribute to the performance enhancement on the backbone model. Nonetheless, the domain-level prompt plays a more significant role on KuaiSAR-small and KuaiSAR dataset, while the user-level prompt is a more essential component on Amazon dataset.

\begin{figure}[t]
\setlength\abovecaptionskip{-0.6\baselineskip}
\setlength\belowcaptionskip{0.3\baselineskip}
	\centering
	\begin{minipage}{0.495\linewidth}
		\centering
        \begin{subfigure}{1\linewidth}
		\includegraphics[width=0.995\linewidth, trim=5 8 5 5,clip]{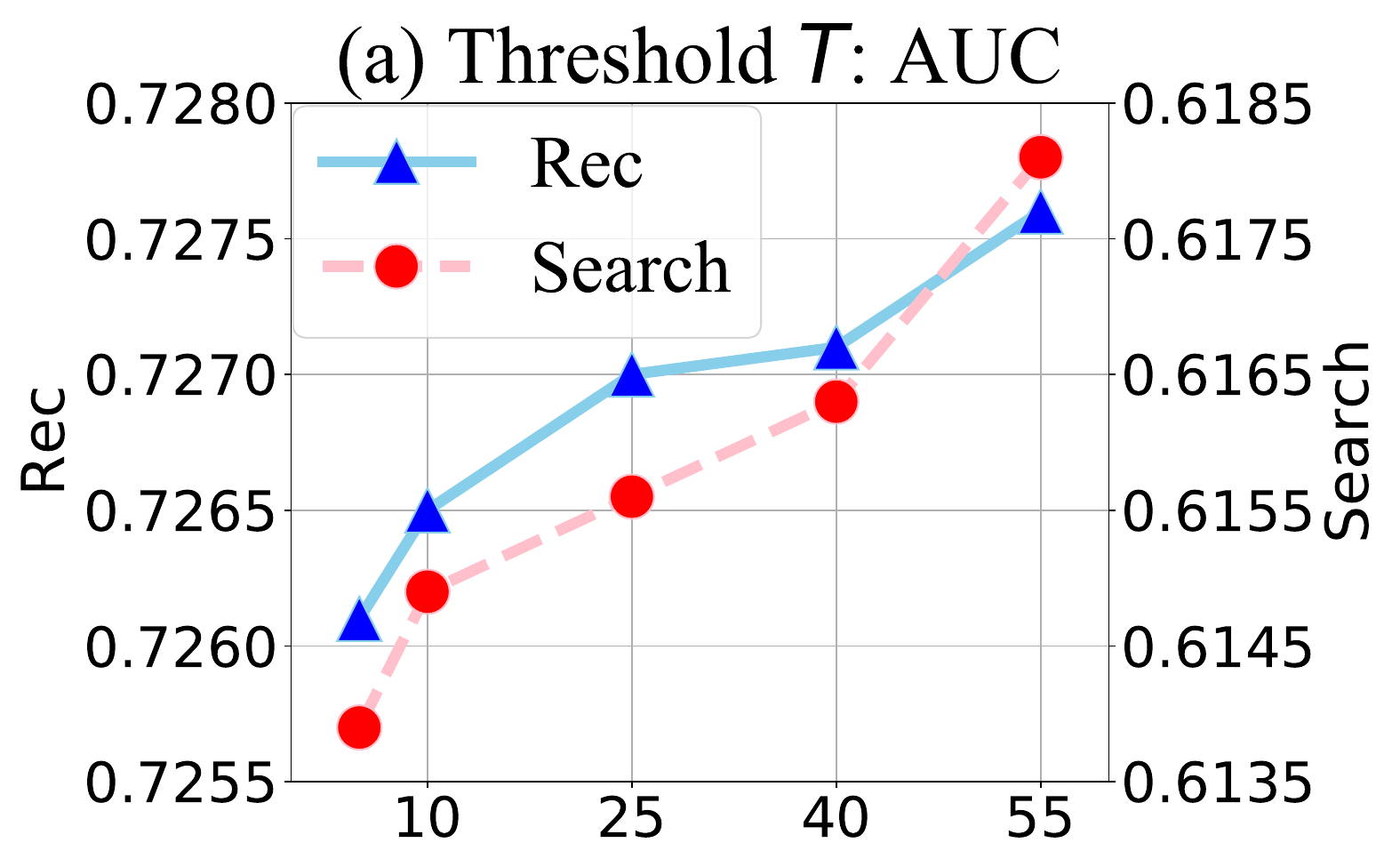}
		\label{fig:RQ3-1}
        \end{subfigure}
	\end{minipage}
	\begin{minipage}{0.495\linewidth}
		\centering
        \begin{subfigure}{1\linewidth}
		\includegraphics[width=0.995\linewidth, trim=5 8 5 5,clip]{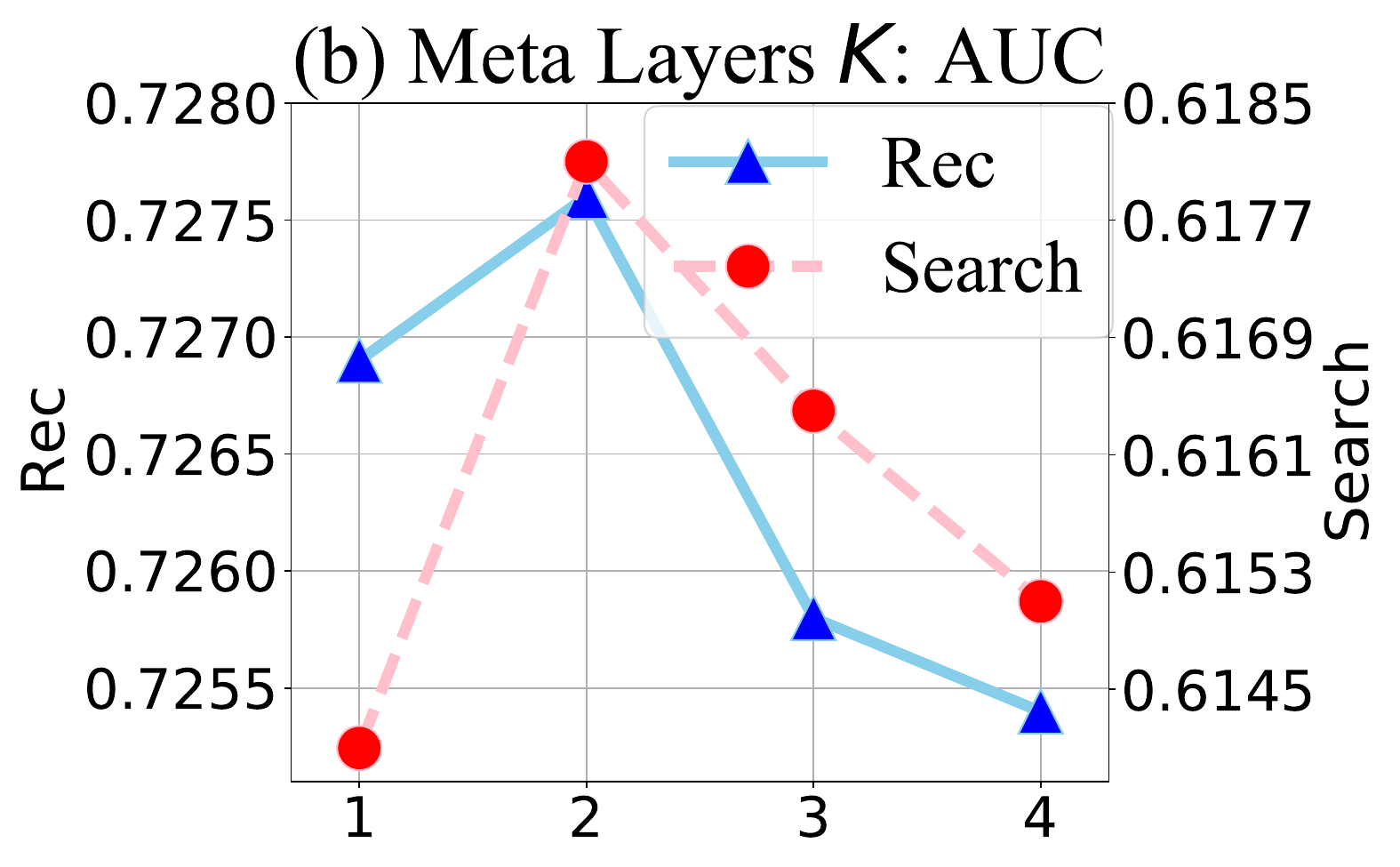}
		\label{fig:RQ3-2}
        \end{subfigure}
	\end{minipage}
	\caption{Analysis on the positive interaction threshold and meta layers evaluated by AUC on KuaiSAR-small dataset.} 
	\label{fig:RQ3}
\end{figure}

On the other hand, comparing `user - bottom' and `user - parallel' we can observe that the former is better while the latter structure even has a negative impact in many cases. Similarly, placing the layers generated from the domain-level prompt in parallel to the scenario tower of the backbone model is the better choice contrasting `domain - bottom' and `domain - parallel'. Consequently, we can draw the conclusion that the combination of layers from user-level prompt at bottom and layers from domain-level prompt in parallel is the optimal architecture. A reasonable explanation is that the user-level multi-scenario information helps personalizing and modulating the general embedding better in a lower level. By contrast, the scenario-level information acts better in a higher level like the auxiliary network adopted in STAR \cite{sheng2021one}. 
Last but not least, it is intuitive to investigate the effect of LLM by replacing the obtained knowledge in the form of hidden states with \textbf{randomly initialized} vectors. However, such enhancement is not observed.

\subsection{Hyper-parameter Analysis (RQ3)}

We analyze the effect of two significant hyper-parameters, i.e., the threshold ($T$) of positive interactions in the user-level prompt and the number of fully connected layers ($K$) generated by the meta network using STAR as the backbone model on KuaiSAR-small dataset.
Specifically, as shown in Figure \ref{fig:RQ3}(a), the recommendation accuracy continuously increases as $T$ rises until the token limit of LLM is reached, since more information about cross-scenario preference of users is introduced and captured by LLM. Besides, from Figure \ref{fig:RQ3}(b) we can see that the best performance is obtained where $K=2$.  Nevertheless, the results degrade as the network becomes deeper and a possible reason is over-fitting.

\subsection{Efficiency Analysis (RQ4)}

\begin{table}[t] \small
\centering
\setlength\abovecaptionskip{0.5\baselineskip}
\setlength\belowcaptionskip{0.2\baselineskip}
\caption{Efficiency comparison on KuaiSAR-small in seconds.}
    \label{efficiency}
    \resizebox{0.475\textwidth}{!}{
    \begin{tabular}{ccccc}
    \toprule
        Models & STAR & STAR\_DN & STAR\_EP & STAR\_ours \\
    \midrule
        Training & 2.8e-2 & 2.9e-2 & 2.9e-2 & 1.5e-1 \\
        Inference & 1.7e-2 & 1.7e-2 & 1.7e-2 & 5.2e-2 \\
    \bottomrule
    \end{tabular}
    }
    \vspace{2.5mm}
\end{table}

To validate the efficiency of our proposed \name paradigm, we take STAR as the backbone model and experiment on KuaiSAR-small dataset. As shown in Table \ref{efficiency}, as an LLM-based method, \name still enables real-time inference even if it requires more training and inference time than other baseline methods.
Besides, the real-world industrial system has a maximum latency specification (let’s say 0.1s), and the present inference time (0.05s in our method) will not undermine the effectiveness of system considering both latency and success rate, while significant improvement of performance is still achieved. Consequently, \name possesses an \textbf{excellent trade-off} in effectiveness and efficiency.

In practice, the last hidden states generated from the user-level prompt can be stored before training, since inferring user preference is the most time-consuming step in \name and it takes about 2 to 4 seconds per user on one single GPU. Besides, the architecture with scenario-level parameters only (i.e., `domain - parallel' in Section \ref{abc}) can serve as an alternative approach because the scenario-level prompt does not require frequent updates or personalization, thus leading to higher efficiency.

\subsection{Case Study (RQ5)} \label{cs}

\begin{figure*}[t]
\centering
\setlength\abovecaptionskip{0.1\baselineskip}
\setlength\belowcaptionskip{-0.1\baselineskip}
\includegraphics[scale=0.515]{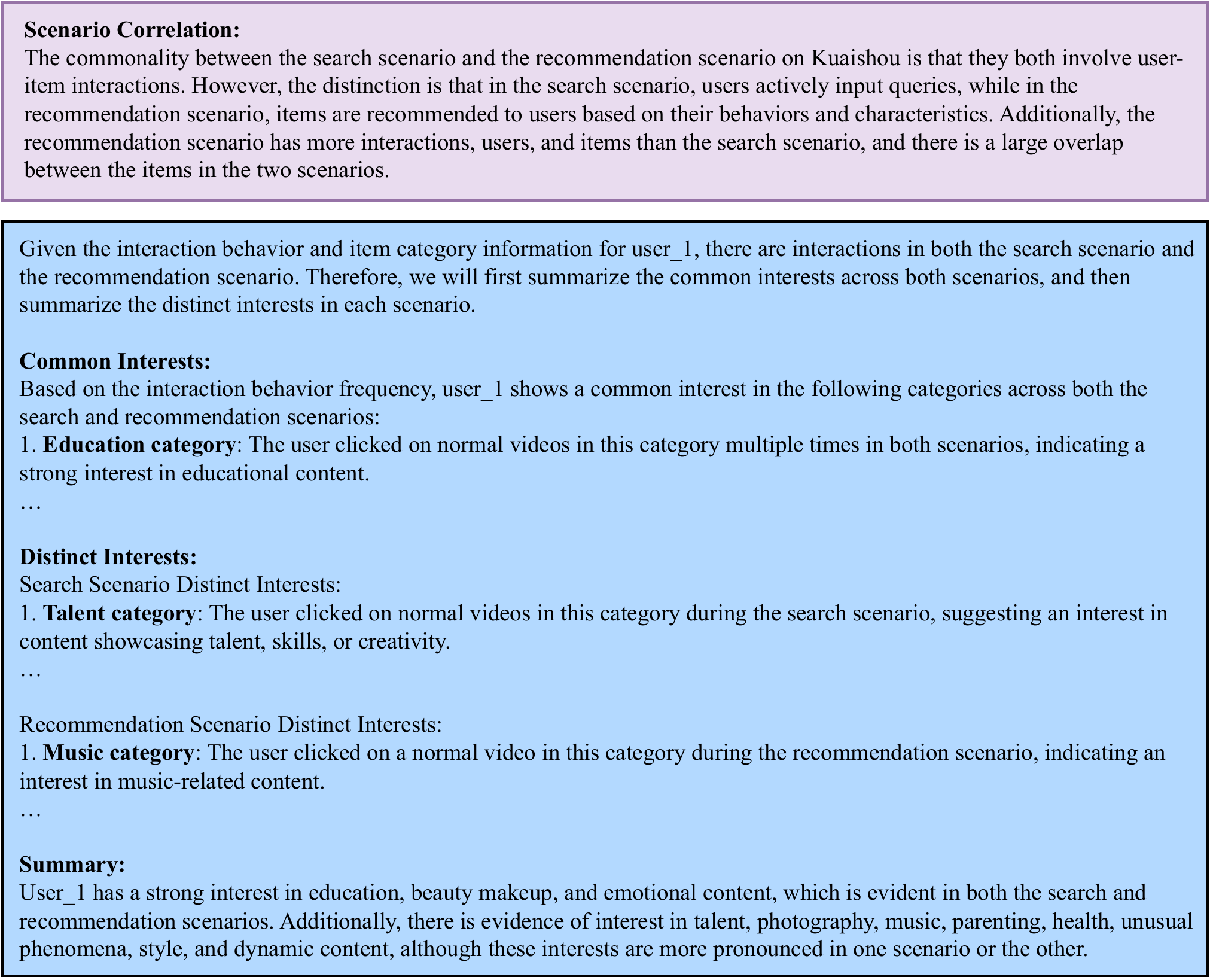}
\caption{Case study of user\_1 on KuaiSAR-small dataset. The purple and blue box denote the result on the scenario correlation and the inferred cross-scenario preferences from scenario- and user-level prompt by ChatGLM2-6B.}
\label{fig:RQ5}
\vspace{-2.5mm}
\end{figure*}

To answer \textbf{RQ5}, we conduct a case study on KuaiSAR-small dataset taking user\_1 as an example and the results generated from the corresponding designed scenario- and user-level prompt by ChatGLM2-6B are shown in Figure \ref{fig:RQ5}.
Specifically, we can see that ChatGLM2-6B not only explicitly captures the commonality and distinction between the recommendation and search scenario, but it also clearly infers user\_1's common and distinct interests among these two scenarios with an accurate user profile summary. Afterward, the generated information is integrated hierarchically by the meta network and meta layers. Consequently, the performance of the multi-scenario backbone model is enhanced. Meanwhile, this knowledge in different levels from LLM also helps improving the interpretability of the original recommender system.

\section{Related Work} \label{rw}

This section brief summarizes the recent work on multi-scenario recommendation and adopting language model for CTR prediction.

\subsection{Multi-Scenario Recommendation}
Multi-scenario recommendation \cite{wang2023plate,wang2022causalint,wang2024diff,li2023hamur}, also known as multi-domain or multi-target cross-domain recommendation \cite{zhu2021cross}, uses the data from all scenarios to simultaneously improve their recommendation performance and tackle data sparsity problem. For example, 
STAR~\cite{sheng2021one} adopts a star-like framework with a shared centered network and scenario-specific networks similar to the parameter sharing pattern design in multi-task modeling \cite{wang2023multi}.
M2M \cite{zhang2022leaving} proposes a meta learning mechanism with hierarchically organized meta attention and meta tower module to grasp inter-scenario correlations and scenario-specific representation.
In addition, to address the imperfectly double seesaw phenomenon, PEPNet \cite{chang2023pepnet} proposes embedding and parameter personalized network to dynamically scale the parameters based on the designed gate mechanism in a domain-specific and task-specific manner, respectively.

Facing the drawbacks of the aforementioned methods introduced in Section \ref{intro}, i.e., insufficient integration of scenario knowledge and neglecting personalized interest across different scenarios, our proposed framework leverages LLM to uncover domain commonality and distinction and infer users' cross-scenario preference, which are further utilized to enhance the multi-scenario backbone model.

\subsection{Language Model for CTR Prediction} 
\label{sec:llm4rec}

Several works try to incorporate Language Models (LMs) including Large Language Models (LLMs) and Pre-trained Language Models (PLMs) 
with recommendation for Click-Through Rate (CTR) prediction \cite{li2023ctrl,fu2023unified,xi2023towards,lin2023clickprompt,wang2023alt,gong2023unified} since they have the potential to bring information gain \cite{zhao2024recommender,xu2023large,li2023e4srec,liu2024moe}. Specifically, one of the solutions is to treat the semantic knowledge as cross-modal information and align with the collaborative signal in conventional recommender systems. For example, CTRL \cite{li2023ctrl} employs contrastive learning for knowledge alignment and integration. KAR \cite{xi2023towards} first generates reasoning knowledge of user preferences and item factual information, then encodes and adapts them to the backbone recommendation models. Besides, S\&R Multi-domain Foundation Model \cite{gong2023unified} simply adopts LLM to extract scenario-invariant text features and tackle item heterogeneity issue.

However, on the one hand, these methods necessitate conducting both training and inference with LM in the sample level, leading to low efficiency of inference and poor deployability on industrial recommender systems. On the other hand, the connection between scenarios and personalized modeling are ignored.
In this paper, we investigate a new LLM-enhanced paradigm, which not only improves the performance of multi-scenario modeling through the inferred scenario-level correlation and user-level cross-scenario interest, but it also enables efficient training and real-time inference. 

\section{Conclusion} \label{conc}
In this paper, we propose an effective efficient interpretable paradigm \name which first leverages LLM to reason about multi-scenario knowledge, including scenario commonality and distinction, and users' cross-scenario preferences through the designed scenario- and user-level prompt. Then it adopts hierarchical meta networks which generate meta layers to adaptively fuse the multi-level knowledge and improve the conventional multi-scenario recommender systems. Extensive experiments on three public datasets validate that \name is not only an effective paradigm compatible with various multi-scenario backbone models, but it is also efficient without fine-tuning LLM and enables real-time recommendation on industrial recommender systems. Besides, the interpretability of conventional recommender system is improved.
%
\appendix

\section{EXPERIMENTAL SETTINGS} \label{detail}
\begin{table}[t] \small
\centering
\setlength\abovecaptionskip{0.1\baselineskip}
\setlength\belowcaptionskip{1\baselineskip}
\caption{The hyper-parameter settings of experiments.}
    \label{hyperdetail}
    \resizebox{0.475\textwidth}{!}{
    \begin{tabular}{llll}
    \toprule
        Dataset & KuaiSAR-small & KuaiSAR & Amazon \\
    \midrule
        Embedding dimension & 16 & 16 & 16 \\
        Batch size & 4096 & 4096 & 1024 \\
        Hidden layers of backbone & [256, 128, 64] & [256, 128, 64] & [64, 32, 16] \\
        Learning rate & 2e-4 & 2e-4 & 2e-3 \\
    \bottomrule
    \end{tabular}
    }
    \vspace{2mm}
\end{table}

The hyper-parameters are summarized in Table~\ref{hyperdetail}. Meanwhile, the designed template of scenario- and user-level prompt are shown in the purple and blue box in Figure \ref{framework}, respectively. Specifically, the most recent 55, 55, and 35 positive interactions, i.e., with click, forward, and like behavior on each scenario are leveraged when constructing the user-level prompt on KuaiSAR-small, KuaiSAR, and Amazon dataset because the maximum context length of ChatGLM2-6B is 8K. Besides, the Adam~\cite{kingma2014adam} optimizer is adopted. The meta network employed is simply a fully connected network.
Overall, 
all the experimental results shown are averaged over 3 runs.

\section{Prompt template} \label{pt}
In this section, we provide the designed scenario-level and user-level prompt template, where the information in square brackets needs to be filled. Specifically, to capture the commonality among scenarios and emphasize distinction for each scenario, the present scenario-level prompts only differ in the instruction: `explicitly summarize the distinction of [scenario\_i name]’, where [scenario\_i name] is the name of each scenario like recommendation or search scenario in KuaiSAR. Meanwhile, other distinctive information of scenario like prior expert knowledge can be provided. That’s the reason why the prompts are constructed separately for each scenario in Algorithm \ref{alg}. 
Prompts with the best performance are:
\vspace{0.5em}
\begin{itemize}[leftmargin=*]

\item \textbf{Scenario-level Prompt:}
\begin{verbatim}
"On platform [dataset name], [platform description], 
suppose there are [specific number] interactions 
in [scenario_1 name], [specific number] interactions 
in [scenario_2 name]... Besides, in these scenarios there 
are [specific number] users [user description] and
[specific number] items [item description] where 
[specific number] users and [specific number] products 
are overlapped. From the relationship among these
scenarios and statistics given, explicitly summarize 
the scenario commonality among the three scenarios
and the distinction of [scenario name]."
\end{verbatim}

\item \textbf{User-level Prompt:}
\begin{verbatim}
"On platform [dataset name], suppose there are
[scenario_1 name], [scenario_2 name]... In [scenario_1 
name], user_[id] [historical behavior + item_type 
+ category]... Besides, in [scenario_2 name], this user 
[historical behavior + item_type + category].
For this user, considering the interaction frequency 
and item title plus description information, explicitly 
summarize the common interest among the three scenarios 
first, then summarize the distinct interest in
each scenario."
\end{verbatim}
\end{itemize}

\section*{ACKNOWLEDGEMENT}
This research was partially supported by Research Impact Fund (No.R1015-23), APRC - CityU New Research Initiatives (No.9610565, Start-up Grant for New Faculty of CityU), CityU - HKIDS Early Career Research Grant (No.9360163), Hong Kong ITC Innovation and Technology Fund Midstream Research Programme for Universities Project (No.ITS/034/22MS), Hong Kong Environmental and Conservation Fund (No. 88/2022), and SIRG - CityU Strategic Interdisciplinary Research Grant (No.7020046), Huawei (Huawei Innovation Research Program), Tencent (CCF-Tencent Open Fund, Tencent Rhino-Bird Focused Research Program), Ant Group (CCF-Ant Research Fund, Ant Group Research Fund), Alibaba (CCF-Alimama Tech Kangaroo Fund (No. 2024002)), CCF-BaiChuan-Ebtech Foundation Model Fund, and Kuaishou.

\bibliographystyle{ACM-Reference-Format}
\balance
\bibliography{ref_new}

\end{document}